\documentclass[12pt]{article}
\usepackage[centertags]{amsmath}
\usepackage{amssymb,amsfonts}
\usepackage[comma,super]{natbib}
\usepackage{latexsym}
\newcommand{\be}{\begin{equation}} \newcommand{\ee}{\end{equation}}
\newcommand{\bea}{\begin{eqnarray}} \newcommand{\eea}{\end{eqnarray}}
\newcommand{\bse}{\begin{subequations}} \newcommand{\ese}{\end{subequations}}
\newcommand{\n}{\nonumber}
\begin{document}
\title{\textbf{Tikekar superdense stars in electric fields}}
\author{K. Komathiraj\thanks{Permanent
address: Department of Mathematical Sciences, South Eastern
University, Sammanthurai, Sri Lanka.}\; and S. D.
Maharaj\thanks{eMail:
\texttt{maharaj@ukzn.ac.za}}\\
Astrophysics and Cosmology Research Unit,\\ School of Mathematical
Sciences, Private Bag X54001,\\ University of KwaZulu-Natal, Durban
4000, South Africa.}
\date{}
\maketitle
\begin{abstract}
\noindent We present exact solutions to the Einstein-Maxwell system
of equations with a specified form of the electric field intensity
by assuming that the hypersurface \{$t$ = constant\} are spheroidal.
The solution of the Einstein-Maxwell system is reduced to a
recurrence relation with variable rational coefficients which can be
solved in general using mathematical induction. New classes of
solutions of linearly independent functions are obtained by
restricting the spheroidal parameter $K$ and the electric field
intensity parameter $\alpha$. Consequently it is possible to find
exact solutions in terms of elementary functions, namely polynomials
and algebraic functions. Our result contains models found previously
including the superdense Tikekar  neutron star model [R. Tikekar,
\emph{J. Math. Phys.} \textbf{31}, 2454 (1990)] when $K=-7$ and
$\alpha=0$. Our class of charged spheroidal models generalise the
uncharged isotropic Maharaj and Leach  solutions
 [S. D. Maharaj and P. G. L. Leach, \emph{J.
Math. Phys.} \textbf{37}, 430 (1996)]. In particular, we find an
explicit relationship directly relating the spheroidal parameter $K$
to the electromagnetic field.
\end{abstract}

\section{Introduction}
Exact solutions of the Einstein-Maxwell field equations are of
crucial importance in  relativistic astrophysics. These solutions
may be utilised to model a charged relativistic star as they  are
matchable to the Reissner-Nordstrom exterior at the boundary. A
recent review of Einstein-Maxwell solutions is given by
Ivanov\cite{Iva}. It is interesting to observe that, in the presence
of charge, the gravitational collapse of a spherically symmetric
distribution of matter to a point singularity may be avoided. In
this situation the gravitational attraction is counterbalanced by
the repulsive Coulombian force in addition to the pressure gradient.
Einstein-Maxwell solutions are also important in studies involving
the cosmic censorship hypothesis and the formation of naked
singularities. The presence of charge affects values for redshifts,
luminosities and maximum mass for stars. Consequently the
Einstein-Maxwell system, for a charged star, has attracted
considerable attention in various investigations.

In an attempt to generate exact solutions in some cases Vaidya and
Tikekar \cite{VaTi} proposed that the geometry of the spacelike
hypersurfaces generated by \{$t$ = constant\} are that of the
3-spheroid. This spheroidal condition provides a clear geometrical
interpretation which is not the case in many other exact solutions.
Knutsen\cite{Kn} was the first to consider the pressure gradients of
stars with spheroidal geometry and showed that they were negative.
Also note that spheroidal geometries exhibit the important physical
feature of being stable with respect to radial pulsations\cite{Knu}.
Tikekar\cite{Tik} comprehensively studied a particular spheroidal
geometry and  showed that it could be applied to superdence neutron
stars with
 densities in the range of $10^{14}~\textrm{g cm}^{-3} $. Maharaj and Leach
 \cite{MaLe} found all spheroidal solutions, for uncharged stars, that could
 be expressed in terms of elementary functions. Mukherjee \emph{et al}\cite{MuPaDa} showed
 that it was possible to express the general solution in terms of
 Gegenbauer functions, an alternate form of the general solution
 was found by Gupta and Jasim\cite{GuJa}. These uncharged solutions can be
 extended to models in the presence of electromagnetic field.
 Spheroidal models in the presence of an electric field have been
 extensively studied by Sharma \emph{et al}\cite{ShMuMa}, Patel and Koppar\cite{PaKo}, Patel \emph{et
 al}\cite{PaTiSa}, Tikekar and Singh\cite{TiSi}, and Gupta and Kumar\cite{GuKu}. These investigations
 have been motivated on the grounds that by restricting the
 geometry of the hypersurfaces \{$t$ = constant\} to be spheroidal
 produces neutral and charged stars which are consistent with
 observations for dense astronomical objects. Models with
 spheroidal geometry can be directly related to particular
 physical situations: the maximum mass is in agreement with values for  cold compact
 objects\cite{ShKaMu}, values for densities are consistent with strange matter\cite{TiJo},
 the equation of state is consistent with a compact X-ray binary
 pulsar Her X-1\cite{ShMu}, relevance to equation of state for stars compared
 of quark-diquark mixtures in equilibrium\cite{Sha}, and uniform charged
 dust in equilibrium\cite{Ti}. Spheroidals geometries are relevant in
 core-envelope steller models, core consisting of isotropic fluid
 and envelope with anisotropic fluid, as shown by Thomas \emph{et
 al}\cite{ThRaVi}
 and Tikekar and Thomas\cite{TiTh}.

It is clear that stars with spheroidal geometry have a number of
different physical applications and therefore require deeper
investigation. In this paper our objective is to generate a class
of charged spheroidal solutions corresponding to a physically
reasonable form for the electric field intensity. Our intention is
to obtain simple forms for the solution that highlights the role
of the spheroidal parameter. In section 2 we express the
Einstein-Maxwell field equations for the static spherically
symmetric line element. The condition of pressure isotropy becomes
a second order linear differential equation which facilitates the
integration process. We assume a solution in a series form which
yields recurrence relations, which we manage to solve from first
principles in section 3. It is then possible to exhibit exact
solutions to the Einstein-Maxwell system. In section 4 we present
polynomials as first solutions and product of polynomials and
algebraic functions as second linearly independent solutions. In
addition we express the general solutions in terms of elementary
functions. We demonstrate that solutions found previously are
special cases of our general treatment. Finally in section 5 we
briefly  discuss the physical viability of our solutions. We
emphasise that our simple approach of utilising the method of
Frobenius for series yields a rich family of Einstein-Maxwell
solutions in terms of elementary functions. This approach was
utilised by Thirukkanesh and Maharaj\cite{ThMa} to produce
electromagnetic solutions to the Einstein-Maxwell system that
contains the Durgapal and Bannerji neutron star model\cite{DuBa}.
\section{The isotropic model}
On physical grounds it is necessary for the gravitational field to
be static and spherically symmetric to describe the internal
structure of a charged relativistic sphere. Therefore we  assume
that the interior of a spherically symmetric star is described by
the line element \be\label{eq:1} ds^{2}
=-e^{2\nu(r)}dt^{2}+e^{2\lambda(r)}dr^{2}+r^{2}(d\theta^{2}+\sin^{2}\theta
d\phi^{2})\ee  in  Schwarzschild coordinates $(t,r,\theta,\phi)$
where $\nu(r)$ and $\lambda(r)$ are arbitrary functions. For
charged perfect fluids the Einstein-Maxwell system of field
equations can be written in the form \bse\label{eq:2}\bea
\frac{1}{r^{2}}(1-e^{-2\lambda})+\frac{2\lambda^\prime}{r}e^{-2\lambda}&=&\rho+\frac{1}{2}E^{2}\\
\label{eq:2.1}-\frac{1}{r^{2}}(1-e^{-2\lambda})+\frac{2\nu^\prime}{r}e^{-2\lambda}&=&p-\frac{1}{2}E^{2}\\
\label{eq:2.2}e^{-2\lambda}\left(\nu^{\prime\prime}+{\nu^\prime}^2+\frac{\nu^\prime}{r}-\nu^\prime\lambda^\prime-\frac{\lambda^\prime}{r}\right)&=&p+\frac{1}{2}E^{2}\\
\sigma&=&\frac{1}{r^{2}}e^{-\lambda}(r^{2}E)^\prime\eea\ese for
the line element (\ref{eq:1}). The field equations (\ref{eq:2})
are the same as those in Thirukkanesh and Maharaj\cite{ThMa}; we
are utilising units in which the coupling constant and the speed
of light are unity. The energy density $\rho$ and the pressure $p$
are measured relative to the comoving fluid 4-velocity
$u^{a}=e^{-\nu}\delta_{0}^{a}$ and primes denote differentiation
with respect to the radial coordinate $r$.  In the system
(\ref{eq:2}), the quantities $E$ and $\sigma$ are the electric
field intensity and the proper charge density respectively.

To integrate the system (\ref{eq:2}) it is necessary to choose two
of the variables $\nu,~\lambda,~\rho,~p$ or $E.$ In our approach we
specify $\lambda$ and $E.$ In the integration procedure we make the
choice \be\label{eq:3}
e^{2\lambda(r)}=\frac{1-Kr^{2}/R^{2}}{1-r^{2}/R^{2}}\ee where $K$ is
an arbitrary constant. The form (\ref{eq:3}) for the gravitational
potential $\lambda$ restricts the geometry of the 3-dimensional
hypersurfaces \{$t$ = constant\} to be spheroidal. When $K=0$ the
hypersurfaces \{$t$ = constant\} become spherical. For the choice
$K=0$ familiar spacetimes are regainable for particulars forms of
the metric function $e^{2\nu}$, e.g. the choice $\nu =0$ gives the
metric of Einstein's universe. On eliminating $p$ from
(\ref{eq:2.1}) and (\ref{eq:2.2}), for the particular form
(\ref{eq:3}), we obtain \bea\label{eq:4}
(1-Kr^{2}/R^{2})^{2}E^{2}&=&(1-
Kr^{2}/R^{2})(1-r^{2}/R^{2})\left(\nu^{\prime\prime}+{\nu^\prime}^{2}-\frac{\nu^\prime}{r}\right)\n\\
&-&(1-K)(r/R^{2})\left(\nu^\prime+\frac{1}{r}\right)+\frac{1-K}{R^{2}}(1-Kr^{2}/R^{2})\eea
which is the condition of pressure isotropy with a nonzero
electromagnetic field.

It is convenient at this point to introduce the transformation
\be\label{eq:5}
\psi(x)=e^{\nu(r)},~~~x^{2}=1-\frac{r^{2}}{R^{2}}\ee Then the
condition of pressure isotropy (\ref{eq:4}) becomes
\be\label{eq:6}
(1-K+Kx^{2})\ddot{\psi}-Kx\dot{\psi}+\left(\frac{(1-K+Kx^{2})^{2}R^{2}E^{2}}{x^{2}-1}+K(K-1)\right)\psi=0\ee
in terms of new variables $\psi$ and $x;$ dots denote
differentiation with respect to $x.$ The Einstein-Maxwell system
(\ref{eq:2}) implies
\bse\label{eq:7}\bea \label{eq:7.1}\rho&=&\frac{1-K}{R^{2}}\frac{(3-K+Kx^{2})}{(1-K+Kx^{2})^{2}}-\frac{1}{2}E^{2}\\
p&=&\frac{1}{R^{2}(1-K+Kx^{2})}\left(-2x\frac{\dot{\psi}}{\psi}+K-1\right)+\frac{1}{2}E^{2}\\
\sigma^{2}&=&\frac{[2xE-(1-x^{2})\dot{E}]^{2}}{R^{2}(1-x^{2})(1-K+Kx^{2})}\eea\ese
in terms of the variable $x.$ Thus $\rho,~p,$ and $\sigma$ are
defined in terms of  $E$ in (\ref{eq:7})

The solution of the Einstein-Maxwell system  depends on the
integrability of (\ref{eq:6}). Clearly (\ref{eq:6}) is integrable
once $E$ is specified. A variety of choices for $E$ is possible;
however only a few are physically reasonable. We need to choose
$E$ such that closed form solutions are possible. We make the
choice \be\label{eq:8} E^{2}=\frac{\alpha
K(x^{2}-1)}{R^{2}(1-K+Kx^{2})^{2}}\ee where $\alpha$ is constant.
A similar form of $E$ was also used by Sharma {\em et al}
\cite{ShMuMa} and Tikekar and Singh \cite{TiSi}. The electric
field intensity $E$ in (\ref{eq:8}) vanishes at the centre of the
star, and remains continuous and bounded in the interior of the
star for a wide range of values of the parameters $\alpha$ and
$K$. Thus this choice for $E$ is physically reasonable and useful
in the study of the gravitational behaviour of charged stars. On
substituting (\ref{eq:8}) into (\ref{eq:6}) we obtain
\be\label{eq:9}(1-K+Kx^{2})\frac{d^{2}\psi}{dx^{2}}-Kx\frac{d\psi}{dx}+K(\alpha+K-1)\psi=0\ee
This is a second order differential equation which is linear in
$\psi.$   We expect that our investigation of (\ref{eq:9}) will
produce viable models of charged stars since the special case
$\alpha=0$ yields models consistent with neutron stars.
\section{ Series solution}
It is possible to express the solution of (\ref{eq:9}) in terms of
special functions namely the Gegenbauer functions as demonstrated by
Sharma {\em et al}\cite{ShMuMa}. However that form of the solution
is not particularly useful because of the analytic complexity of the
special functions involved. In addition the role of parameters of
physical interest, such as the spheroidal parameter $K,$ is lost or
obscured in the representation as Gegenbauer functions. The
representation of the solutions in a simple form is necessary for a
detailed physical analysis. Consequently we attempt to obtain a
general solution to the differential equation (\ref{eq:9}) in a
series form using the method of Frobenius. Later we will indicate
that it is possible to extract solutions in terms of polynomials and
algebraic functions for particular parameter values.

As the point $x=0$ is a regular point of (\ref{eq:9}), there exist
two linearly independent solutions of the form of a power series
with centre $x=0.$ Thus we assume
\be\label{eq:10}\psi(x)=\sum_{i=0}^{\infty}a_{i}x^{i}\ee where the
constants $a_{i}$ are the coefficients of the series. For a
legitimate solution we need to determine the coefficients $a_{i}$
explicitly. On substituting (\ref{eq:10}) into (\ref{eq:9}) we
obtain after simplification
 \bea\label{eq:10a}
 (1-K)2.1a_{2}+K(\alpha+K-1)a_{0}+[(1-K)3.2a_{3}+K(\alpha+K-2)a_{1}]x\n\\
 +\sum_{i=2}^{\infty}\{(1-K)(i+1)(i+2)a_{i+2}+K[\alpha+K-1+i(i-2)]a_{i}\}x^{i}=0\eea
in increasing powers of $x.$ For equation (\ref{eq:10a}) to be
valid for all $x$ in the interval of convergence we require
\label{eq:11}\bse\bea
\label{eq:11.1}(1-K)2.1a_{2}+K(\alpha+K-1)a_{0}&=&0\\
\label{eq:11.2}(1-K)3.2a_{3}+K(\alpha+K-2)a_{1}&=&0\\
\label{eq:11.3}(1-K)(i+1)(i+2)a_{i+2}+K[\alpha+K-1+i(i-2)]a_{i}&=&0,~i\geq2\eea\ese
Equation (\ref{eq:11.3}) is the the linear recurrence relation
governing the structure of the solution.

The recurrence relation (\ref{eq:11.3}) consists of variable,
rational coefficients. It does not fall in the known class of
difference equations and has to be solved from first principles.
It is possible to solve (\ref{eq:11.3}) using the principle of
mathematical induction. We first consider the even coefficients
$a_{0},~a_{2},~a_{4},~.~. ~.$  Equation (\ref{eq:11.1}) implies
\be\label{eq:11a}
a_{2.1}=\left(\frac{K}{K-1}\right)^{1}\frac{1}{(2.1)!}\prod_{q=1}^{1}[\alpha+K-1+(2q-2)(2q-4)]a_{0}\ee
where we have utilised the conventional symbol $\prod$ to denote
multiplication for the first term. We now assume the
pattern\be\label{eq:11b}
a_{2p}=\left(\frac{K}{K-1}\right)^{p}\frac{1}{(2p)!}\prod_{q=1}^{p}[\alpha+K-1+(2q-2)(2q-4)]a_{0}\ee
for the coefficient $a_{2p}$ which is the inductive step. We now
establish that this is true for the next coefficient $a_{2(p+1)}.$
Replacing $i$ with $2p$ in (\ref{eq:11.3}) we obtain
\bea\label{eq:11c}
a_{2(p+1)}&=&\left(\frac{K}{K-1}\right)\left(\frac{\alpha+K-1+2p(2p-2)}{(2p+2)(2p+1)}\right)a_{2p}\n\\
&=&\left(\frac{K}{K-1}\right)\left(\frac{\alpha+K-1+2p(2p-2)}{(2p+2)(2p+1)}\right)\times\n\\
&&\left(\frac{K}{K-1}\right)^{p}\frac{1}{(2p)!}\prod_{q=1}^{p}[\alpha+K-1+(2q-2)(2q-4)]a_{0}\n\\
&=&\left(\frac{K}{K-1}\right)^{p+1}\frac{1}{[2(p+1)]!}\prod_{q=1}^{p+1}[\alpha+K-1+(2q-2)(2q-4)]a_{0}
\eea where we have used (\ref{eq:11b}). Hence by mathematical
induction all the even coefficients $a_{2i}$ can be written in
terms of the coefficient $a_{0}.$ These coefficients generate a
pattern \be\label{eq:12}
a_{2i}=\left(\frac{K}{K-1}\right)^{i}\frac{1}{(2i)!}\prod_{q=1}^{i}[\alpha+K-1+(2q-2)(2q-4)]a_{0}\ee
for the even coefficients $a_{0},~a_{2},~a_{4},~.~.~.$

We can obtain a similar formula for the odd coefficients $a_{1},~
a_{3},~a_{5},~.~.~.$  From (\ref{eq:11.2}) we have
\be\label{eq:12a}
a_{2.1+1}=\left(\frac{K}{K-1}\right)^{1}\frac{1}{(2.1+1)!}\prod_{q=1}^{1}[\alpha+K-1+(2q-1)(2q-3)]a_{1}\ee
for the first term. We now assume \be\label{eq:12b}
a_{2p+1}=\left(\frac{K}{K-1}\right)^{p}\frac{1}{(2p+1)!}\prod_{q=1}^{p}[\alpha+K-1+(2q-1)(2q-3)]a_{1}\ee
for the coefficient $a_{2p+1}$. We then establish that this is
true for the next coefficient $a_{2(p+1)+1}$. Replacing $i$ with
$(2p+1)$ in (\ref{eq:11.3}) we obtain \bea\label{eq:12c}
a_{2(p+1)+1}&=&\left(\frac{K}{K-1}\right)\left(\frac{\alpha+K-1+(2p+1)(2p-1)}{(2p+3)(2p+2)}\right)a_{2p+1}\n\\
&=&\left(\frac{K}{K-1}\right)\left(\frac{\alpha+K-1+(2p+1)(2p-1)}{(2p+3)(2p+2)}\right)\times\n\\
&&\left(\frac{K}{K-1}\right)^{p}\frac{1}{(2p+1)!}\prod_{q=1}^{p}[\alpha+K-1+(2q-1)(2q-3)]a_{1}\n\\
&=&\left(\frac{K}{K-1}\right)^{p+1}\frac{1}{[2(p+1)+1]!}\prod_{q=1}^{p+1}[\alpha+K-1+(2q-1)(2q-3)]a_{1}\eea
on utilising (\ref{eq:12b}). Hence by using mathematical induction
all the odd coefficients $a_{2i+1}$ can be written in terms of the
coefficient $a_{1}.$ These coefficients generate a pattern which
is clearly of the form \be\label{eq:13}
a_{2i+1}=\left(\frac{K}{K-1}\right)^{i}\frac{1}{(2i+1)!}\prod_{q=1}^{i}[\alpha+K-1+(2q-1)(2q-3)]a_{1}\ee
for the odd coefficients $a_{1},~a_{3},~a_{5},~.~.~.$

The coefficients $a_{2i}$ are generated from (\ref{eq:12}). The
coefficients $a_{2i+1}$ are generated from (\ref{eq:13}). Hence
the difference equation (\ref{eq:11.3}) has been solved and all
nonzero coefficients are expressible in terms of the leading
coefficients $a_{0}$ and $a_{1}$. From (\ref{eq:10}),
(\ref{eq:12}) and (\ref{eq:13}) we establish that
\bea \label{eq:14}\psi(x)&=&a_{0}\left(1+\sum_{i=1}^{\infty}\left(\frac{K}{K-1}\right)^{i}\frac{1}{(2i)!}\prod_{q=1}^{i}[\alpha+K-1+(2q-2)(2q-4)]x^{2i}\right)\n\\
&+&a_{1}\left(x+\sum_{i=1}^{\infty}\left(\frac{K}{K-1}\right)^{i}\frac{1}{(2i+1)!}\prod_{q=1}^{i}[\alpha+K-1+(2q-1)(2q-3)]x^{2i+1}\right)\n\\\eea
where $a_{0}$ and $a_{1}$ are arbitrary constants. Clearly
(\ref{eq:14}) is of the form \be
\psi(x)=a_{0}\psi_{1}(x)+a_{1}\psi_{2}(x)\ee where\bse
\label{eq:14.1}\bea
\label{eq:14.2}\psi_{1}(x)&=&\left(1+\sum_{i=1}^{\infty}\left(\frac{K}{K-1}\right)^{i}\frac{1}{(2i)!}\prod_{q=1}^{i}[\alpha+K-1+(2q-2)(2q-4)]x^{2i}\right)\\
\label{eq:14.3}\psi_{2}(x)&=&\left(x+\sum_{i=1}^{\infty}\left(\frac{K}{K-1}\right)^{i}\frac{1}{(2i+1)!}\prod_{q=1}^{i}[\alpha+K-1+(2q-1)(2q-3)]x^{2i+1}\right)\n\\\eea\ese
are linearly independent solutions of (\ref{eq:9}). Thus we have
found the general series solution to the differential equation
(\ref{eq:9}) for the choice of the electromagnetic field $E$ given
in (\ref{eq:8}). The solution (\ref{eq:14}) is expressed in terms
of a series with real arguments unlike the complex arguments given
by software packages. The series (\ref{eq:14.2}) and
(\ref{eq:14.3}) converge if there exist a nonnegative value for
the radius of convergence. Note that the radius of convergence of
the series is not less than the distance from the centre $(x=0)$
to the nearest root of the leading coefficient of the differential
equation (\ref{eq:9}). Clearly this is possible for a wide range
of values for $K$.
\section{Solutions with elementary functions}
It is interesting to observe that the series in (\ref{eq:14.1})
terminates for restricted values of the parameters $\alpha$ and
$K$. This will happen when $\alpha+K$ takes on specific integer
values. Utilising this feature it is possible to generate
solutions in terms of elementary functions by determining the
specific restriction on $\alpha$ and $K$ for a terminating series.
Solutions in terms of polynomials and algebraic functions can be
found. We use the recurrence relation (\ref{eq:11.3}), rather than
the series (\ref{eq:14.1}), to find the elementary solutions as
this is simpler.
\subsection{Polynomial solutions}
We first consider polynomials of even degree. It is convenient to
set \bse \bea i&=&2(j-1)\\
K+\alpha&=&2-(2n-1)^{2}\eea \ese where $n>1$ is fixed integer in
(\ref{eq:11.3}). This leads to \be
\label{eq:16}a_{2j}=-\gamma\frac{(n+j-2)(n-j+1)}{2j(2j-1)}a_{2j-2}\ee
where we have set $\gamma=4-\frac{4}{4n(n-1)+\alpha}.$ We note
that (\ref{eq:16}) implies $a_{2(n+1)}=0$. Consequently the
remaining coefficients $a_{2(n+2)},~a_{2(n+3)},~.~.~.$ vanish.
Equation (\ref{eq:16}) may be solved to yield \be\label{eq:17}
a_{2j}=(-\gamma)^{j}\frac{(n+j-2)!}{(n-j)!(2j)!},~~0\leq j\leq
n\ee where we have set $ a_{0}=\frac{1}{n(n-1)}$. With the help of
 (\ref{eq:17}) we can express the polynomial in even powers of $x$ in
the form
\be\label{eq:18}f_{1}(x)=\sum_{j=0}^{n}(-\gamma)^{j}\frac{(n+j-2)!}{(n-j)!(2j)!}x^{2j}\ee
for $K+\alpha=2-(2n-1)^{2}.$

We now consider polynomials of odd degree. For this case we let
\bse \bea i&=&2(j-1)+1\\
K+\alpha&=&2(1-2n^{2})\eea \ese where $n>0$ is fixed integer in
(\ref{eq:11.3}). We obtain \be\label{eq:19}
a_{2j+1}=-\mu\frac{(n+j-1)(n-j+1)}{2j(2j+1)}a_{2j-1}\ee where we
have set $\mu=4-\frac{4}{(4n^{2}-1+\alpha)}.$ We observe that
(\ref{eq:19}) implies $a_{2(n+1)+1}=0$. Consequently the remaining
coefficients $a_{2(n+2)+1},~a_{2(n+3)+1},~.~.~.$ vanish. Equation
(\ref{eq:19}) can be solved to yield \be\label{eq:20}
a_{2j+1}=(-\mu)^{j}\frac{(n+j-1)!}{(n-j)!(2j+1)!},~~0\leq j\leq n
\ee where we have set $a_{1}=\frac{1}{n}$. With the assistance of
(\ref{eq:20}) we can express the polynomial in odd powers of $x$
as \be\label{eq:21}
g_{1}(x)=\sum_{j=0}^{n}(-\mu)^{j}\frac{(n+j-1)!}{(n-j)!(2j+1)!}x^{2j+1}\ee
for $ K+\alpha=2(1-2n^{2})$.

The polynomial solutions (\ref{eq:18}) and (\ref{eq:21}) comprise
the first solution of (\ref{eq:9}) for appropriate values of
$K+\alpha$.
\subsection{Algebraic solutions} We take the second solution of (\ref{eq:9}) to be of the form  \be
\psi(x)=u(x)(1-K+Kx^{2})^{3/2}\ee when $u(x)$ is an arbitrary
polynomial. Particular solutions found in the past are special
cases of this general form; the factor $(1-K+Kx^{2})^{3/2}$ helps
to simplify the integration process. This motivates the algebraic
form for $\psi$ as a generic solution to the differential equation
(\ref{eq:9}). On substituting $\psi$ in (\ref{eq:9}) we obtain
after simplification
\be\label{eq:22}(1-K+Kx^{2})\frac{d^{2}u}{dx^{2}}+5Kx\frac{du}{dx}+K(\alpha+K+2)u=0\ee
which is linear differential equation for $u(x)$.

As in \S 4.1 we can find two classes of polynomial solutions for
$u(x)$, in even powers of $x$ and in odd powers of $x$, for
certain values of $K+\alpha.$ As the point $x=0$ is a regular
point of (\ref{eq:22}), there exists two linearly independent
solutions of the form of power series with centre $x=0$. Therefore
we can write \be\label{eq:23}
u(x)=\sum_{i=0}^{\infty}b_{i}x^{i}\ee where $b_{i}$ are the
coefficients of the series. Substituting (\ref{eq:23}) in
(\ref{eq:22}) we obtain \bea\label{eq:23a}
(1-K)2.1b_{2}+K(\alpha+K+2)b_{0}+[(1-K)3.2b_{3}+K(\alpha+K+7)b_{1}]x\n\\
+\sum_{i=2}^{\infty}\{(1-K)(i+2)(i+1)b_{i+2}+K[\alpha+K+2+i(i+4)]b_{i}\}x^{i}=0\eea
For equation (\ref{eq:23a}) to hold true for all $x$ we require
that \bse\bea
(1-K)2.1b_{2}+K(\alpha+K+2)b_{0}&=&0\\
(1-K)3.2b_{3}+K(\alpha+K+7)b_{1}&=&0\\
\label{eq:23.1}(1-K)(i+2)(i+1)b_{i+2}+K[\alpha+K+2+i(i+4)]b_{i}&=&0
,i\geq2 \eea\ese which governs the coefficients.

We first consider even powers of $x$. Replacing
 $i$ with $2(j-1)$ and assuming  $K+\alpha=2(1-2n^{2})$ in
(\ref{eq:23.1}), where $n>0$ is fixed integer, we obtain
\be\label{eq:24} b_{2j}=-\mu\frac{(n+j)(n-j)}{2j(2j-1)}b_{2j-2}\ee
where we have set $\mu=4-\frac{4}{4n^{2}-1+\alpha}$. From
(\ref{eq:24}) we have that $b_{2n}=0$ and  subsequent coefficients
$b_{2(n+1)},~b_{2(n+2)},~.~.~.$ vanish. Then  (\ref{eq:24}) has
the solution \be\label{eq:25}
b_{2j}=(-\mu)^{j}\frac{(n+j)!}{(n-j-1)!(2j)!},~0\leq j\leq n-1\ee
where we have set $b_{0}=n$. On using (\ref{eq:23}) and
(\ref{eq:25}) the polynomial in even powers of $x$ leads to the
expression  \be \label{eq:26}
g_{2}(x)=(1-K+Kx^{2})^{3/2}\sum_{j=0}^{n-1}(-\mu)^{j}\frac{(n+j)!}{(n-j-1)!(2j)!}x^{2j}\ee
for $K+\alpha=2(1-2n^{2})$.

We now consider odd powers of $x$. Replacing  $i$ with $2(j-1)+1$
and assuming  $K+\alpha=2-(2n-1)^{2}$ in (\ref{eq:23.1}), where
$n>1$ is fixed integer, we obtain \be \label{eq:27}
b_{2j+1}=-\gamma\frac{(n+j)(n-j-1)}{2j(2j+1)}b_{2j-1}\ee where we
have set $\gamma=4-\frac{4}{4n(n-1)+\alpha}$. From (\ref{eq:27})
we have that  $b_{2(n-1)+1}=0$ and  subsequent coefficients
$b_{2n+1},~b_{2(n+1)+1},~.~.~.$ vanish. Then equation
(\ref{eq:27}) has the solution \be\label{eq:28}
b_{2j+1}=(-\gamma)^{j}\frac{(n+j)!}{(n-j-2)!(2j+1)!},~0\leq j\leq
n-2\ee where we have set $b_{1}=n(n-1).$ On using (\ref{eq:23})
and (\ref{eq:28}) the polynomial in odd powers of $x$ leads to the
result \bea \label{eq:29}
f_{2}(x)&=&(1-K+Kx^{2})^{3/2}\sum_{j=0}^{n-2}(-\gamma)^{j}\frac{(n+j)!}{(n-j-2)!(2j+1)!}x^{2j+1}\eea
for $K+\alpha=2-(2n-1)^{2}$.

The algebraic solutions (\ref{eq:26}) and (\ref{eq:29}) comprise the
second solution of (\ref{eq:9}) for appropriate values of
$K+\alpha$. The solutions (\ref{eq:26}) and (\ref{eq:29}) are
expressed as products of algebraic functions and polynomials, and
they are clearly linearly independent from (\ref{eq:21}) and
(\ref{eq:18}), respectively.
\subsection{ Elementary functions} We have
obtained two classes of polynomial solutions (\ref{eq:18}) and
(\ref{eq:21}) in \S 4.1 to the differential equation (\ref{eq:9}).
Also we have found two classes of algebraic solutions (\ref{eq:26})
and (\ref{eq:29}) in \S 4.2. By collecting these results we can
express the general solution to (\ref{eq:9}) in two categories. The
first category of solution for $\psi(x)= f(x)$ is given
by \bea\label{eq:30} f(x)&=&A f_{1}(x)+B f_{2}(x)\n\\\n\\
&=&A\sum_{j=0}^{n}(-\gamma)^{j}\frac{(n+j-2)!}{(n-j)!(2j)!}x^{2j}\n\\
&+&B(1-K+Kx^{2})^{3/2}\sum_{j=0}^{n-2}(-\gamma)^{j}\frac{(n+j)!}{(n-j-2)!(2j+1)!}x^{2j+1}\eea
for the values\bse\bea\gamma&=&4-\frac{4}{4n(n-1)+\alpha}\\
K+\alpha&=&2-(2n-1)^{2}.\eea\ese The second category of solution for
$\psi(x)= g(x)$
has the form \bea\label{eq:31}g(x)&=&A g_{1}(x)+B g_{2}(x)\n\\\n\\
&=&A\sum_{j=0}^{n}(-\mu)^{j}\frac{(n+j-1)!}{(n-j)!(2j+1)!}x^{2j+1}\n\\
&+&B(1-K+Kx^{2})^{3/2}\sum_{j=0}^{n-1}(-\mu)^{j}\frac{(n+j)!}{(n-j-1)!(2j)!}x^{2j}\eea
for the values\bse\bea\mu&=&4-\frac{4}{4n^{2}-1+\alpha}\\
K+\alpha&=&2(1-2n^{2})\eea \ese where $A$ and $B$ are arbitrary
constants.

It is remarkable that these solutions are expressed completely as
combinations of polynomials and algebraic functions. From our
general class of solutions  (\ref{eq:30}) and (\ref{eq:31}) it is
possible to generate particular solutions found previously.
Consider one example with $\alpha=0$ and $K=-7(n=2).$ Then
$\gamma=\frac{7}{2}$ and it is easy to verify that equation
(\ref{eq:30}) becomes \be
\psi=A\left(1-\frac{7}{2}x^{2}+\frac{49}{24}x^{4}\right)+Bx\left(1-\frac{7}{8}x^{2}\right)^{\frac{3}{2}}.\ee
Thus we have regained the Tikekar\cite{Tik} solution for a
superdense neutron star from our general solutions. Many other
particular solutions found in the literature are also contained in
our general solutions, eg. the model of Patel and
Koppar\cite{PaKo}. The solutions (\ref{eq:30}) and (\ref{eq:31})
reduce to the Maharaj and Leach\cite{MaLe} model when $\alpha=0$.
Our solutions are applicable to a charged superdense star with
spheroidal geometry. When $\alpha=0$ we obtain uncharged
relativistic stars which model ultradense barotropic matter.
\section{Physical conditions}
In the general solution of Sharma \emph{et al}\cite{ShMuMa} it is
not possible to isolate the spheroidal parameter $K$ as that
solution is given in terms of special functions. Our solutions are
in terms of simple elementary functions which facilitate a study
of the physical features, in particular the role of $K$. The exact
solutions (\ref{eq:30}) and (\ref{eq:31}) make it possible to
analyse the role of the spheroidal parameter $K$ and it's
connection to the electromagnetic field. In particular it is
possible to  make the following comment about the special role
that the spheroidal parameter $K$ has in charged solutions. The
form of the solution for the uncharged relativistic star is
similar to (\ref{eq:30}) and (\ref{eq:31}); however the models are
different because the coefficients of the polynomials (namely
$\gamma$ and $\mu$) differ by the parameter $\alpha.$ If the
parameter $\alpha\geq0$ then we observe that \be
K_{(\alpha\neq0)}<K_{(\alpha=0)}\n\ee Hence the presence of charge
directly affects the spheroidal geometry through the parameter
$K.$ The geometry of the hypersurfaces \{$t$ = constant\} in the
spacetime manifold is related to the electromagnetic field via the
relationships $K=2-(2n-1)^{2}-\alpha$ and $K=2(1-2n^{2})-\alpha.$
Such explicit relationships connecting the spacetime geometry to
the energy momentum (or electromagnetic field) are rare in exact
solutions. The presence of charge $\alpha$ decreases the value of
the spheroidal parameter $K$ in our solutions.

We make a few brief comments about the physics of the models found
in this paper. If $0<K<1(\alpha<0)$ then $\rho$ remains positive
in the region \be
(1-x^{2})<\frac{3(1-K)}{K(1-K-\alpha/2)}\Longrightarrow
r^{2}<\frac{3R^{2}(1-K)}{K(1-K-\alpha/2)}\ee which restricts the
size of the configuration. When $K<0(\alpha>0)$ there is no
restriction on $\rho$. Hence $\rho$ is positive in the interior of
the star. It is clear from (\ref{eq:7.1}) and (\ref{eq:8}) that
$d\rho/dr<0$ for $K<0(\alpha>0)$. Consequently the energy density
decreases from the centre to the boundary. For the pressure to be
vanish at the boundary $r=a$ we require that \be
\left(\frac{2}{R^{2}\sqrt{1-a^{2}/R^{2}}}\left[\frac{\dot{\psi}}{\psi}\right]_{r=a}-\frac{1}{a^{2}}\right)\frac{1-a^{2}/R^{2}}{1-Ka^{2}/R^{2}}+\frac{1}{a^{2}}+\frac{\alpha
Ka^{2}/R^{2}}{2R^{2}(1-Ka^{2}/R^{2})^{2}}=0\ee where $\psi$ is
given by (\ref{eq:30}) or (\ref{eq:31}). This will constrain the
values of the constants $A$ and $B$. The solution of the
Einstein-Maxwell equations for $r>a$ is given by the
Reissner-Nordstrom metric as \be\label{eq:3m}
ds^{2}=-\left(1-\frac{2m}{r}+\frac{q^{2}}{r^{2}}\right)dt^{2}+\left(1-\frac{2m}{r}+\frac{q^{2}}{r^{2}}\right)^{-1}dr^{2}+r^{2}(d\theta^{2}+\sin^{2}\theta
d\phi^{2}), \ee where $m$ and $q$ are the total mass and charge of
the star respectively. To match the line element (\ref{eq:1}) with
the Reissner-Nordstrom metric (\ref{eq:3m}) across the boundary at
$r=a$ we require the continuity of the gravitational potentials
and of the radial electric field at $r=a$. Continuity of the
gravitational potentials yields the relationships between the
constants $A, B, K, a$ and $R$ as \bse\bea
\left(1-\frac{2m}{a}+\frac{q^{2}(a)}{a^{2}}\right)&=&[A\psi_{1}(a)+B\psi_{2}(a)]^{2}\\
\left(1-\frac{2m}{a}+\frac{q^{2}(a)}{a^{2}}\right)^{-1}&=&\frac{1-Ka^{2}/R^{2}}{1-a^{2}/R^{2}}\eea\ese
The continuity of electric field yields the form \be
q^{2}(a)=-\frac{\alpha Ka^{6}/R^{4}}{(1-Ka^{2}/R^{2})^{2}}\ee for
the charge at the boundary. This shows that continuity of the
metric functions across the boundary $r=a$ is easily achieved. The
matching conditions at $r=a$ may place restrictions on the metric
coefficients $\nu$ and its first derivative for uncharged matter;
and the pressure may be nonzero if there is a surface layer of
charge. However there are sufficient free parameters to satisfy
the necessary conditions that arises for a particular spheroidal
model. It is interesting to note that our solutions may be
interpreted as models for relativistic anisotropic stars where the
parameter $\alpha$ plays a role of the anisotropy factor.
Isotropic and uncharged stars can be regained when $\alpha=0$.
Chaisi and Maharaj\cite{ChMa}, Dev and Gleiser\cite{DeGl, DeGle}
and Maharaj and Chaisi\cite{MaCh} provide some recent treatments
involving the physics of
anisotropic matter. \\

\noindent{\large \bf Acknowledgements}\\

\noindent KK thanks the National Research Foundation and the
University of KwaZulu-Natal for financial support, and also extends
his appreciation to the South Eastern University of Sri Lanka for
granting study leave. SDM acknowledges that this work is based upon
research supported by the South African Research Chair Initiative of
the Department of Science and Technology and the National Research
Foundation. We are grateful to the referee for a careful  reading of
the manuscript.
%
%
%----------------bibliography----------------------------------
%\bibliographystyle{phaip}
%\bibliography{biblio,papersbib}
%---------------end of bibliography----------------------------
%%%%%%%%%%%%%%%%%%%%%%%%%%%%%%%%%%%%%%%%%%%%%%%%%%%%%%%%%%%%%%

%%%%%%%%%%%%%%%%%%%%%%%%%%%%%%%%%%%%%%%%%%%%%%%%%%%%%%%%%%%%%%%
%
\end{document}